\documentclass[doublecol]{epl2} 
\usepackage{graphicx}
\usepackage{rotating}
\usepackage{amssymb}    
\usepackage{amsmath}   
\usepackage{epsfig}
\usepackage[normalem]{ulem}   
\usepackage{bm}   
\usepackage{color}
\date{\today}

\title{Coupling-induced periodic windows in networked discrete-time systems}

\author{Leonard Hallier\inst{1} \and Everton S. Medeiros\inst{1,2} \and Antonio Mihara\inst{3} \and Rene O. Medrano-T\inst{3,4} \and Anna Zakharova\inst{1}}

\institute{                    
  \inst{1} Institut f\"ur Theoretische Physik, Technische Universit\"at Berlin, Hardenbergstra\ss e 36, 10623 Berlin, Germany\\
  \inst{2} Institute for Chemistry and Biology of the Marine Environment, Carl von Ossietzky University Oldenburg, 26111 Oldenburg, Germany\\
  \inst{3} Departamento de Física, Universidade Federal de São Paulo,UNIFESP, 09913-030, Campus Diadema, São Paulo, Brasil\\
  \inst{4} Departamento de Física, Instituto de Geociências e Ciências Exatas, Universidade Estadual Paulista, UNESP, 13506-900, Campus Rio Claro, São Paulo, Brasil
}

\abstract{Networked nonlinear systems present a variety of emergent phenomena as a result of the mutual interactions between their units. An interesting feature of these systems is the presence of stable periodic behavior even when each unit oscillates chaotically if in isolation. Surprisingly, the mechanism in which the network interaction replaces chaos by periodicity is still poorly understood. Here, we show that such an onset of regularity can occur via replication of periodic windows. This phenomenon multiplies the stability domains in the system parameter space, not only suppressing chaos but also making the network less vulnerable to external disturbances such as shocks and noise. Moreover, we observe that the network cluster synchronizes for the parameters corresponding to the replica periodic windows. To confirm these observations, we employ the formalism of the master stability function demonstrating that the complete synchronized state is indeed transversally unstable in the replica windows.}

\begin{document}

\maketitle

In the last two decades, many efforts have been devoted to quantifying the stability of networks composed of dynamical units. In this direction, reliable approaches to assess both linear \cite{Pecora1998,Pecora2000,Mihara2019} and global \cite{Girvan2006, Menck2013} stability of synchronized states are available. However, the stability of the synchronized behavior is not the only concern, for the sake of the network's plain functioning, additional attention is required to harness possible unstable behavior. 

In particular, the occurrence of chaotic behavior in many applications is one type of instability that threatens the system's functioning with unpredictability and large solution variability. In this scenario, several approaches have been developed to control chaos in low-dimensional systems \cite{Ott1990, Braiman1991,Pyragas1992, Boccaletti2000} and some of them were also applied to high-dimensional systems \cite{Scholl2008,Chacon2016,Chacon2017}. Interestingly, often low-dimensional chaotic systems spontaneously exhibit stable periodic behavior when networked, e.g., coupled with identical copies of themselves \cite{Omelchenko2012,Dziubak2013,Semenova2015,Zur2018}. In addition, networks of coupled discrete-time systems have been investigated in the context of synchronization \cite{Pikovsky2001, Bauer2009, Boccaletti2018}, partial synchronization patterns \cite{Zur2018, Rybalova2019, Zakharova2020, Zhang2021}, switching in time between different types of partial synchronization patterns \cite{Semenova2017} or sub-patterns \cite{Zhang2020}. However, much less attention has been paid to the investigation of complex periodic windows in such networks. Moreover, despite of the general knowledge about the bifurcations giving rise to periodicity, the mechanisms in which the mutual interactions within the network establish stable periodic behavior in a collection of chaotic systems are not completely understood.      

To shed light on these mechanisms, one may consider the occurrence of mutual disturbances among the network units as a starting point. From this perspective, the concept of non-feedback methods for suppressing chaos can be applicable. In general, these techniques are characterized by external interventions in the system dynamics aiming to suppress chaotic behavior without destroying the system's main characteristics. In particular, for periodically driven systems, a very effective non-feedback method to eliminate chaotic behavior is the addition of a second periodic forcing with a small amplitude to the system dynamics \cite{Lima1990,Meucci1994,Chacon1995,Pisarchik1997,Mirus1999,Zambrano2006,Meucci2016,HegedHus2020,Martinez2020}. Remarkably, this extra periodic forcing has been found not only to mitigate chaos for individual sets of system parameters but for large extensions of the system parameter space \cite{Medeiros2010,Medeiros2011}. Specifically, the weak periodic forcing replicates intervals of the system parameter space containing parameters leading to stable periodic behavior, i.e., periodic windows. Generally, the replica windows contain stable periodic orbits with the same period and stability characteristics as the ones populating the original window. This phenomenon has been called {\it replication of periodic windows} \cite{Medeiros2010,Medeiros2011}.

Following these developments, the multiplication of periodic windows has been observed also for discrete-time systems in which an external time-alternating perturbation creates identical copies of pre-existing periodic windows \cite{Manchein2017}. Interestingly, in this study, the period of the perturbation corresponds to a multiplication factor to the number of periodic windows. The system exhibits duplication, triplication, and quadruplication of periodic windows when the perturbation is of periods two, three, and four, respectively \cite{Manchein2017}. Later on, the same phenomenon has been proposed as a mechanism to increase the availability of stable domains and, consequently, reduce the effects of noise and parameter inaccuracies in dynamical systems such as the Hénon map, Langevin equation, and Chua's circuit \cite{Silva2018}. Moreover, replication of periodic windows has been recently shown to occur in a model for two asymmetrically coupled neurons \cite{Santana2021}. Coupling asymmetries have been also demonstrated to suppress chaotic behavior in larger oscillator networks \cite{Medeiros2021} whereas unit asymmetries (heterogeneity) can also induce synchronous states \cite{Nishikawa2016,Zhang2017,Molnar2020}.

Here, we report that the replication of periodic windows is an intrinsic mechanism for the onset of periodicity in symmetrical networks of identical discrete-time units. More specifically, we demonstrate that the mutual interaction among the units replaces the periodic perturbations and, the replication of periodic windows occurs in these systems without external interventions. We first show how the distance between the replica windows depends on the coupling intensity among the units. Next, we investigate the interplay between the synchronized behavior of the network and the replication of periodic windows. We observe that original periodic windows correspond to states of complete synchronization in the network, while in replica windows, the network contains phase cluster states. These states are characterized by network clusters containing completely synchronized units oscillating out-of-phase with other clusters. Remarkably, we observe replica periodic windows in which the number of phase clusters equals the total number of units in the system, this solution corresponds to traveling waves in the network. Finally, we implement the formalism of the master stability function to demonstrate that completely synchronized states are transversally unstable for the parameters composing the replica periodic windows. This confirms the general occurrence of the phase clusters.

\section{Network Model}

\begin{figure}[!htp]
\centering
\includegraphics[width=8.5cm,height=2.75cm]{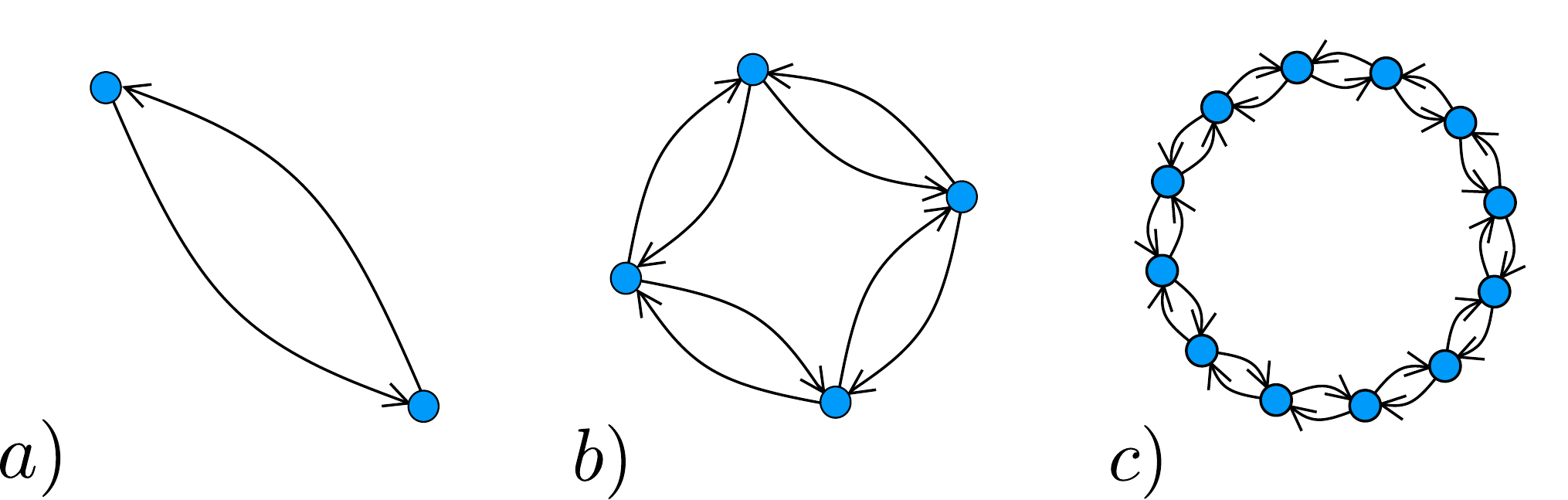}
\caption{Schematic of the coupling structure for different network sizes. (a) $N=2$. (b) $N=4$. (c) $N=12$.}
\label{figure_1}
\end{figure}

We consider a network of identical one-dimensional discrete-time units diffusively coupled via their defining function. Each unit is coupled to its first neighbors and periodic boundary conditions are adopted. The dynamical equation describing this system is given by:
\begin{equation}
    \theta_{i}^{t+1}  = f(\theta_i^t) + \frac{\sigma}{D_i}\sum_{j\in\mathcal{D}_i}[f(\theta_{j}^{t})-f(\theta_{i}^{t})] \;\;\;\;\;\; \text{mod(1)},
    \label{Circle}
\end{equation}
where the variable $\theta_{i}^{t}$ defines the dynamical state of the unit $i$ with $i=1,\dots,N$ at a discrete-time $t$. The constants $\sigma$ and $N$ control the coupling strength among the maps and the network size, respectively. The set $\mathcal{D}_i$ prescribes the adjacency of a unit $i$, while the constant $D_i$ is the respective number of units connected to $i$. The local dynamics of the network is given by the Circle map $f(\theta)=\theta+\Omega-\frac{K}{2\pi}\sin(2\pi\theta)$. The behavior of this particular map is determined by two control parameters, defining a two-dimensional parameter space $K \times \Omega$. This parameter plane contains the periodic windows of interest in this work, constituting the sole reason for our choice of this particular system. In Fig. \ref{figure_1}, we show a schematic of the coupling structures approached through the text.

\section{Coupling-induced Replications}

\begin{figure}[!htp]
\centering
\includegraphics[width=8.5cm,height=3.7cm]{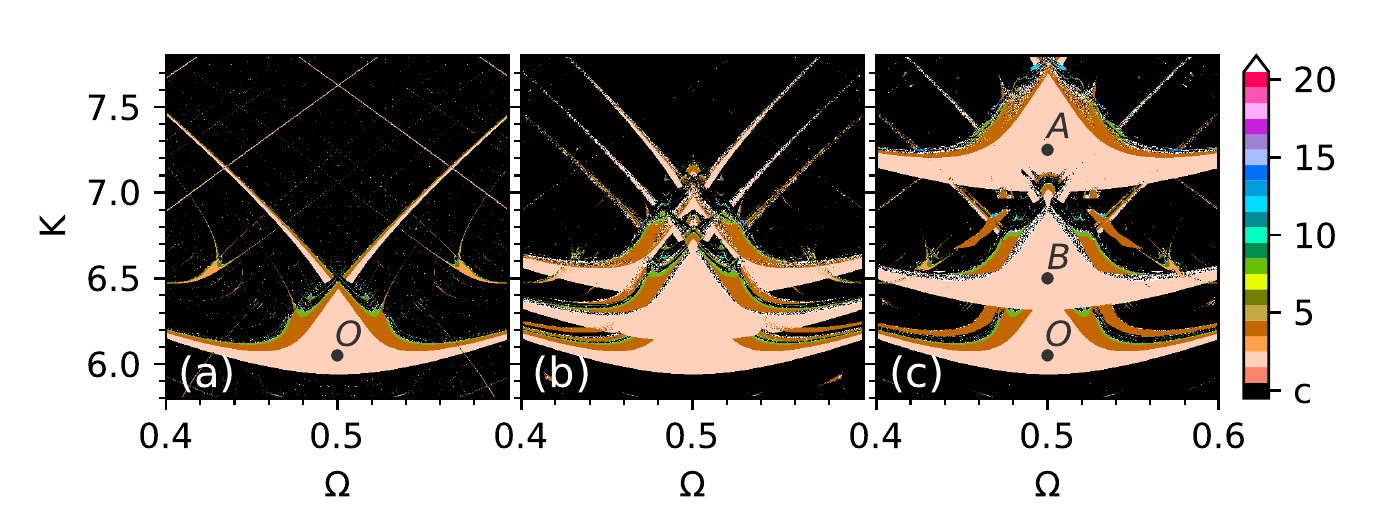}
\caption{Parameter space $K \times \Omega$ of the network containing $N=4$ units. The color code stands for the period $p$ of stable periodic orbits. (a) Periodic window $O$ for the uncoupled system ($\sigma=0$). (b) For $\sigma=0.125$, the original $O$ and two replica periodic windows $A$ and $B$ overlap. (c) For $\sigma=0.22$, the three windows are further apart from each other.}
\label{figure_2}
\end{figure}

To effectively demonstrate the replication of periodic windows in the parameter space of the network, we first examine the parameter space of the uncoupled system, i.e., $\sigma=0$ in Eq. (\ref{Circle}). For that, in Fig. \ref{figure_2}(a), we color code the period $p$ (period-$p$) of orbits occurring in a subset of the parameter plane $K \times \Omega$ of the map. The parameter pairs corresponding to orbits with $p\leq20$ are marked with different colors (see color bar), the points corresponding to $p>20$ are marked in gray, while black corresponds to chaotic behavior. The periodic window $O$ in Fig. \ref{figure_2}(a) is a typical structure of periodicity in the parameter space of nonlinear dynamical systems known by different names such as crossroad area \cite{Carcasses1991}, shrimp-shaped window \cite{Gallas1993}, compound window \cite{Lorenz2008}, among others \cite{Fraser1982,Markus1989}. These windows have been observed in the parameter space of low-dimensional mathematical models of a number of systems \cite{Bonatto2008,Gallas2010,Oliveira2011,Stegemann2011,Celestino2011,Freire2014,Stegemann2014,Santos2016,Costa2016,Varga2020,Raphaldini2021} and in laboratory experiments with electronic circuits \cite{Baptista2008,Stoop2010,Viana2010}. The bifurcations delimitating these windows are well understood. Typically, a saddle-node bifurcation (boundary between the black and apricot regions in Fig. \ref{figure_2}(a)) gives rise to the main period-$2$ orbit in the window. In turn, this orbit undergoes period-doubling, creating subharmonic period-$4$ (brown region) orbits which underlie the next segment of the periodic window. The successive period-doubling bifurcations develop the periodic orbits into a chaotic attractor (upper boundary in Fig. \ref{figure_2}(a)).

We now consider the coupled system with $N=4$ as in the schematic shown in Fig. \ref{figure_1}(b). For $\sigma=0.125$ in Fig. \ref{figure_2}(b), we observe that, due to the network coupling, the original periodic window is multiplied by three. With this, a substantial portion of the parameter plane $K \times \Omega$ corresponding to chaotic behavior at the level of each isolated map leads to stable periodic behavior in the network. Interestingly, the two new replicated windows possess the same period-$2$ main periodic orbit as the one previously existing in the uncoupled map. In contrast, the subharmonic orbits present in the original window are not entirely preserved in the replicas, they appear intermingled with irregular solutions. For instance, notice in the brown regions in Fig. \ref{figure_2}(b) and Fig. \ref{figure_2}(c) that the windows corresponding to subharmonic orbits are most affected at their borders, coinciding with regimes of slow convergence to the periodic orbits. These parameter regions are prone to the onset of additional attractors due to the network coupling. The coexistence of the subharmonic orbits with irregular solutions in these regions gives their observed riddled aspect. Next, in Fig. \ref{figure_2}(c), by increasing the coupling strength among the network units, for $\sigma=0.22$ we observe that the periodic windows move apart from each other. This ability offers a level of controllability to the network, making possible the suppression of chaos in pre-determined regions of the parameter space or tuning the network to exhibit periodic trajectories with the desired period.

\begin{figure}[!htp]
\centering
\includegraphics[width=8.4cm,height=3.5cm]{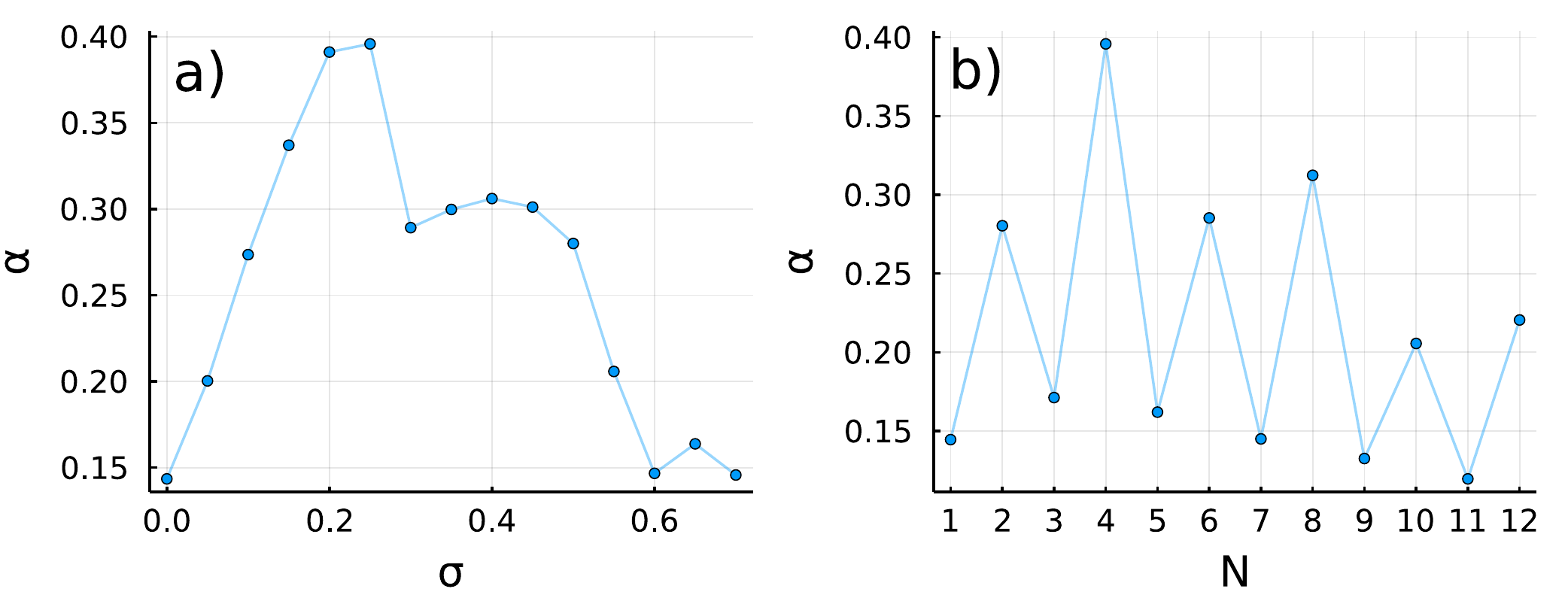}
\caption{(a) The regularity index $\alpha$ as a function of the coupling strength $\sigma$. The network size fixed at $N=4$. (b) The regularity index $\alpha$ as a function of the network size $N$. The coupling intensity is fixed at $\sigma=0.25$}
\label{figure_3}
\end{figure}

The results in Fig. \ref{figure_2} suggest the replication of periodic windows as a mechanism for the appearance of stable periodic behavior in networked systems. However, the replication occurs locally in the system parameter space for windows containing orbits with a given period. In different regions of the parameter space, the network coupling may as well replace periodic windows of the uncoupled system by irregular behavior. This raises the question about the net effect of the network coupling in a large portion of the parameter space. To shed light on this issue, we estimate the intervals of the $K \times \Omega$ plane leading to stable periodic orbits. For that, we define a finite region of the parameter space as $\mathcal{E} = \{ (\Omega,K) \in \mathbb{R}^2 \mid \Omega \in [0.4,0.6], K \in [5.8,7.8] \}$. The pairs ($\Omega$, $K$) resulting in stable periodic orbits are identified by their largest Lyapunov exponent $\mathbf{\Lambda}_{1}<0$, and denoted by $\mathcal{S}$. Then, we estimate the regularity-index $\alpha$ as the intersection of $\mathcal{S}$ and $\mathcal{E}$:
\begin{equation}
 \alpha=\frac{Vol(\mathcal{S} \cap \mathcal{E})}{Vol(\mathcal{E})}.
\end{equation}
With this, we investigate the local abundance of stable periodic orbits in the system's parameter space by computing $\alpha$ as a function of the network parameters. First, in Fig. \ref{figure_3}(a), we obtain $\alpha$ as function of the coupling strength $\sigma$. For a network with $N=4$ units, we observe a regime of growth in the regularity-index $\alpha$ corresponding to the replication of the period-$2$ window shown in Fig. \ref{figure_2}. After the measure $\alpha$ reaches a local maximum around $\sigma \approx 0.22$, it abruptly decreases for values of $\sigma$ corresponding to the quick collapse of the replica windows $A$. For further increasing $\sigma$, there is a second local maximum corresponding to the separation of the window $B$ from the original $O$. Subsequently, the regularity index decreases to the levels of the uncoupled system due to the disappearance of the window $B$. Next, to gather more insights on the stability gains, in Fig. \ref{figure_3}(b), we compute $\alpha$ for different network sizes $N$. The most striking characteristic of these results is the strong influence of the parity of $N$ on $\alpha$. The regularity index is significantly higher for networks of even sizes, this behavior is already a clue to the mechanism underlying the replications. Specifically, we demonstrate later that the formation of phase cluster synchronization is essential for the phenomenon. In special, for the investigated parameter region, networks with an even number of units favor the stability of periodic orbits with even periods due to the possibility of symmetric distributions of units in the attractors. Considering the period-$2$ orbit, which dominates the periodicity region (apricot color in Fig. \ref{figure_2}), for an even $N$, the phase cluster state is composed of two clusters with $N/2$ units in each part of the attractor. For an odd $N$ this symmetry is broken and the stability is no longer favored. This parity preference is due to the balance between the coupling in networks with even sizes and the attractors of the local dynamics with even periods. One can see in Fig. \ref{figure_3}(b) that for odd network sizes, $N=7$ and $N=9$, the stability gain decreases significantly. Another consequence of the symmetric organization of the units in the attractors extension, is the optimal network size $N=4$ for the stability gain seen in Fig. \ref{figure_3}(b). This occurs because, for the specific coupling intensity considered in this figure, the network size $N=4$ is the configuration providing the largest number of different phase clusters. Finally, the regularity index decreases for larger $N$, we attribute this decline in the stability gain to the onset of phenomena in higher dimensions that generally difficult the occurrence of periodic windows, e.g., multistability, hyperchaos, and long-lasting transients in large networks \cite{Medeiros2018, Medeiros2019}.

\begin{figure}[!htp]
\centering
\includegraphics[width=8.5cm,height=4.5cm]{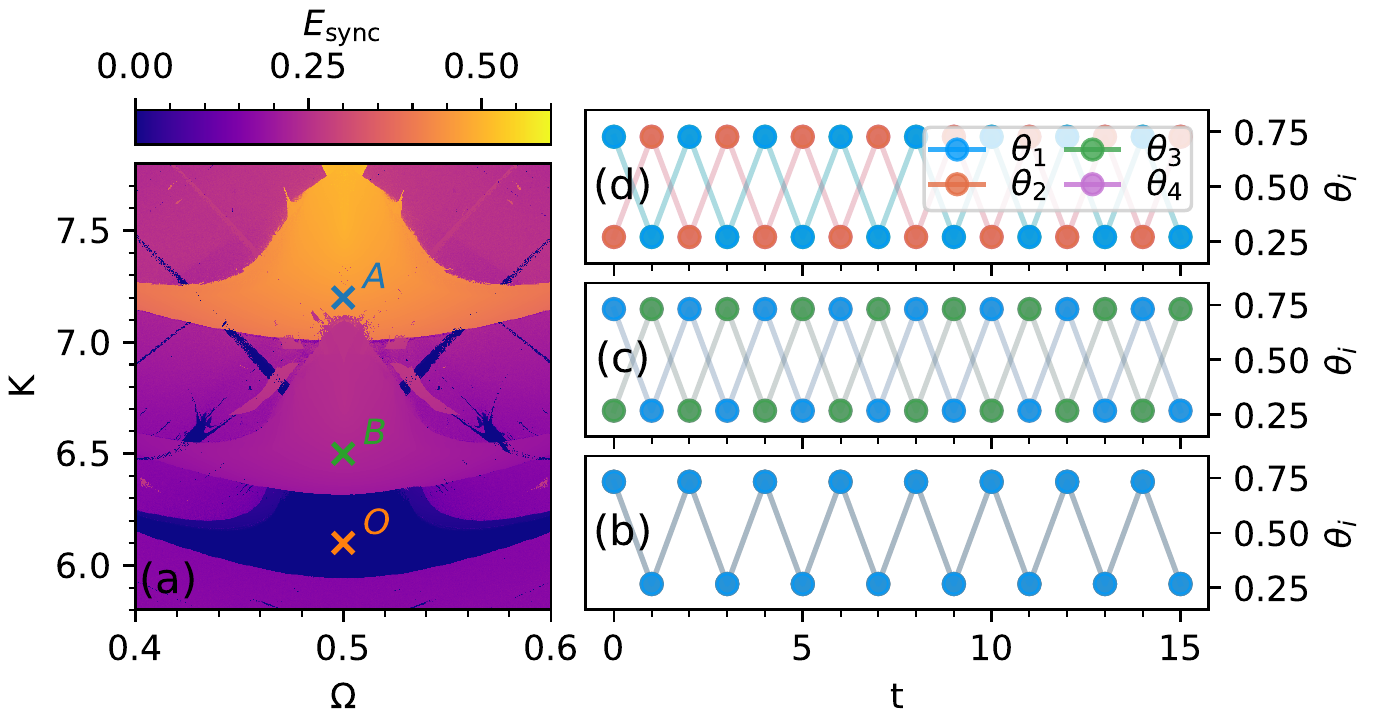}
\caption{For the network parameters fixed at $N=4$ and $\sigma=0.22$. (a) Synchronization error $E_\text{sync}$ (color code) as a function of $K$ and $\Omega$. For $\Omega=0.5$ in (b) Time evolution for the complete synchronized network state in the original window $O$, $K=6.1$.  (c) Time evolution of the phase clusters in the window $B$, $K=6.5$. (d) Time evolution of the phase clusters in the window $A$, $K=7.2$.}
\label{figure_4}
\end{figure}

Since the networks addressed in this study are composed of identical units, one possible solution for these systems is complete synchronization. For this behavior, each map in the network evolves identically in time. As a consequence, the coupling term in Eq. (\ref{Circle}) vanishes and the maps exhibit the solution of the uncoupled case. Therefore, the occurrence of completely synchronized states is incompatible with the replacement of chaotic orbits by stable periodic ones. Hence, next, we unravel the interplay between synchronization and the replication of periodic windows in the network parameter space. First, we introduce a global synchronization measure to detect synchronous behavior among the units as:
\begin{equation}
	E_\text{sync}
	= \frac{1}{NT} \sum_{t=t_0}^{t_0+T} \sum_{i=1}^{N}\left|\theta_{i+1}^{t}-\theta_{i}^{t}\right|,
	\label{sync_error}
\end{equation}
computed over a time interval of length $T=500$ after a transient is disregarded. In Fig. \ref{figure_4}(a), for $N=4$ and $\sigma=0.22$, we obtain the synchronization error $E_\text{sync}$ as a function of the parameters $K$ and $\Omega$, for which the replication of periodic windows takes place in the parameter space (same parameters as in Fig. \ref{figure_2}(c)). We observe that the parameters composing the original periodic window $O$ corresponds to $E_\text{sync}=0$, i.e., the complete synchronized solution (dark blue in Fig. \ref{figure_4}(a)). In Fig. \ref{figure_4}(b), we show the time evolution for a parameter set from the original periodic window illustrating the complete synchronized behavior, $\theta_{1}^{t}=\theta_{2}^{t}=\theta_{3}^{t}=\theta_{4}^{t}$ around the period-$2$ orbit. Conversely, in Fig. \ref{figure_4}(a), we observe that the two replica windows, $A$ and $B$, corresponds to network solutions with $E_\text{sync}\neq 0$. Therefore, inside these windows, the network is synchronized in frequency, but not in phase. In addition, the values of $E_\text{sync}$ are close to uniform inside the respective windows, indicating the occurrence of only one synchronization pattern within the window. However, the values of $E_\text{sync}$ differ among the replica windows, i.e., synchronization patterns are not the same for different replica windows. In fact, by obtaining the time evolution for a parameter set in window $B$, in Fig. \ref{figure_4}(c), we observe the formation of two phase clusters containing two completely synchronized units around the period-$2$ orbit, i.e., $\theta_{1}^{t}=\theta_{2}^{t}$, $\theta_{3}^{t}=\theta_{4}^{t}$ and $\theta_{2}^{t}\neq\theta_{3}^{t}$. For a parameter set in window $A$, two different phase clusters are formed $\theta_{1}^{t}=\theta_{3}^{t}$, $\theta_{2}^{t}=\theta_{4}^{t}$, and with $\theta_{2}^{t}\neq\theta_{3}^{t}$ around the same period-$2$ orbit, producing a larger synchronization error. In addition, for $\sigma\rightarrow 0$, all three network synchronization patterns discussed in Fig. \ref{figure_4} coexist in the region of the original periodic window.

\begin{figure}[!htp]
\centering
\includegraphics[width=8.6cm,height=5.7cm]{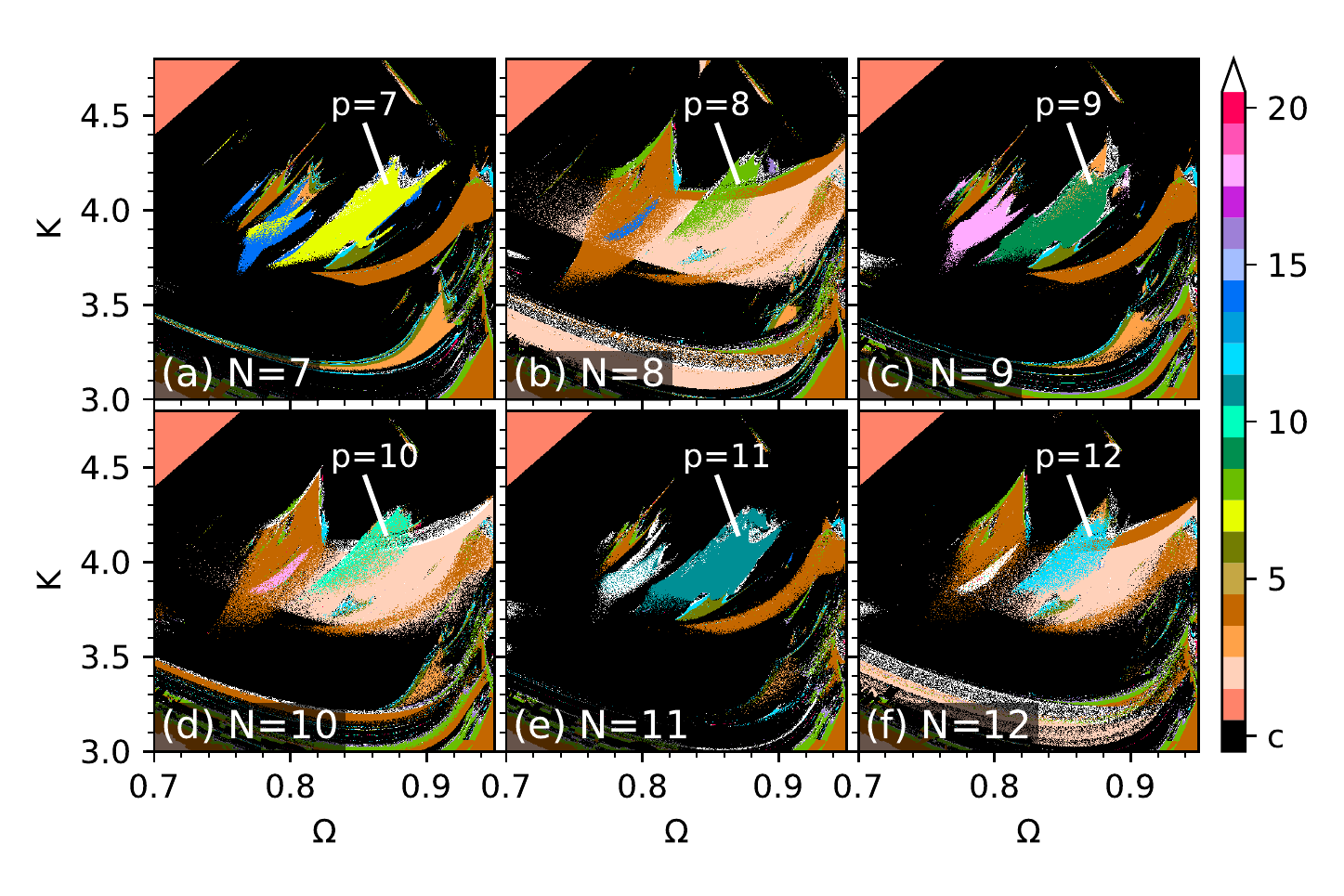}
\caption{Parameter space $K \times \Omega$ for different network sizes $N$. The periodic windows are color-coded following their respective period. The windows identified by their period $p$ correspond to the ones, in which $p=N$, giving rise to traveling waves in the network. The coupling intensity is fixed at $\sigma=0.25$.}
\label{figure_5}
\end{figure}

For larger networks (as in Fig. \ref{figure_1}(c)), the number of phase clusters in the replica windows will vary and produce a diversity of solutions. For $p<N$, the synchronized states are clustered in different phases of the periodic orbit, for $p>N$ not all possible phases available in the periodic orbit are occupied. One interesting scenario emerges when the number of different phase clusters equals the network size $p=N$, i.e., all units are out of phase around the same periodic orbit. This configuration results in replica windows containing waves traveling along the network as a solution. We illustrate this behavior for different network sizes in Figs. \ref{figure_5}(a)-\ref{figure_5}(f). In these figures, the windows containing the traveling waves are marked with the respective period of their orbits. The periods match the size of the respective network sizes.

\begin{figure}[!htp]
\centering
\includegraphics[width=8cm,height=8cm]{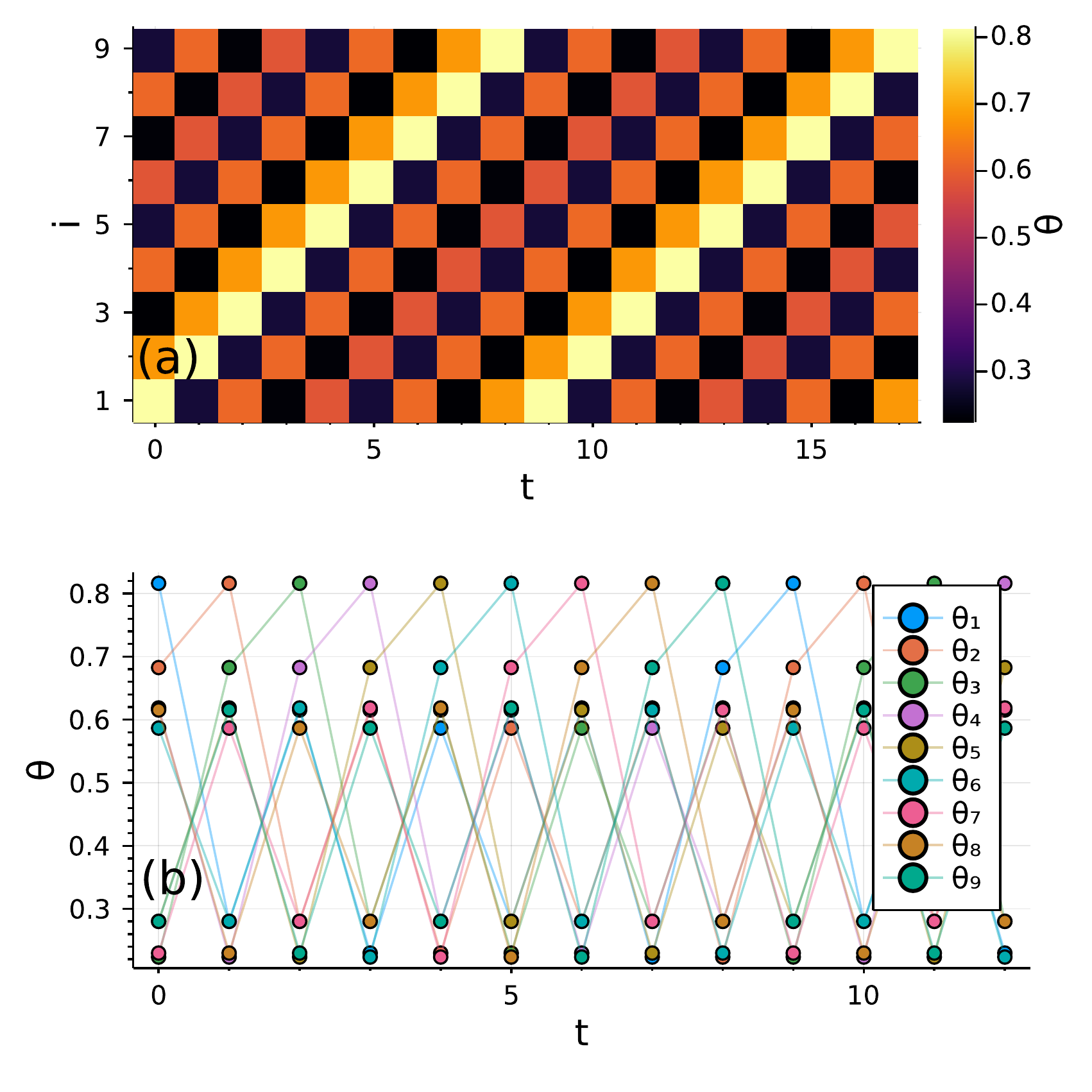}
\caption{(a) The space-time plot showing the traveling wave occurring inside a replica window for $\Omega=0.87$, $K=4.15$, $N=9$, and $\sigma=0.25$ (see Fig. \ref{figure_5}). (b) Time evolution of each network unit composing the traveling wave. The $N=9$ units follow the same period-$9$ periodic orbit occupying $9$ different phases.}
\label{figure_6}
\end{figure}

In Fig. \ref{figure_6}(a), for $N=9$ (Fig. \ref{figure_5}(e)), we obtain a space-time plot to show the corresponding traveling wave solution in the network. In Fig. \ref{figure_6}(b), we show the time evolution of all units in the network exhibiting different phases in a period-$9$ orbit, i.e., $\theta_{1}^{t}\neq \dots \neq \theta_{9}^{t}$. Note that some units possess very similar phases, which indicates small deviations in the phases are preventing complete synchronization and, consequently, chaotic oscillations.    

In this section, we observe that the appearance of replica periodic windows is associated with phase cluster states of the network. Therefore, one may conclude that the onset of such windows requires the absence of completely synchronized states in the network. In the next section, we confirm this possibility by obtaining the transversal stability of complete synchronized states.

\section{Mechanism}

Now, we demonstrate the connection between complete synchronization being transversally unstable and the onset of replica periodic windows in the network. For that, we employ the formalism of the master stability function (MSF) for networked discrete-time systems. We first write the dynamical equation describing the network in a more general form:
\begin{eqnarray}
 \theta_{i}^{(t+1)} &=&  f(\theta_i^t)+\sigma \sum_{j=1}^{N}G_{ij}h(\theta_j^t) \;\;\;\;\;\; \text{mod(1)}.
 \label{Circle2}
\end{eqnarray}
Different from the original system defined in Eq. (\ref{Circle}), in which the maps are diffusively coupled via their defining function $f(\theta^t)$, in Eq. (\ref{Circle2}), the maps are coupled by a linear coupling function $h(\theta^t)$. This facilitates the implementation of MSF and it also exemplifies the occurrence of our reported phenomenon in a different network setup. The coupling matrix ${\bf G}$ satisfies the condition $\sum_{j=0}^{N}G_{ij}=0$ for any $i$ and it can be diagonalized with a set of eigenvalues $\{\lambda_i,i=1,\dots,N\}$. Perturbing the solution $\theta^t$ of Eq. (\ref{Circle2}) as $\delta \theta^t\equiv \theta^t-s(t)$, we obtain the following variational:
\begin{eqnarray}
 \delta \theta_i^{(t+1)} &=& Df(s)\cdot\delta \theta_i^t +\sigma\sum_{j=1}^{N}G_{ij}Dh(s)\cdot\delta\theta_j^t,
 \label{Circle3}
\end{eqnarray}
where $Df(s)$ and $Dh(s)$ are the respective derivatives of $f$ and $h$ evaluated at $s$. Applying block diagonalization to the second term of Eq. (\ref{Circle3}), we obtain the equation decoupled in modes of ${\bf G}$:
\begin{eqnarray}
 \delta \xi_i^{(t+1)} &=& \left[Df(s)+\sigma \lambda_i Dh(s) \right]\cdot\delta \xi_i^t \;\;\;\;\;\; \text{mod(1)}.
 \label{diagonalized}
\end{eqnarray}
Substituting $\sigma \lambda_i=\gamma$ in Eq. (\ref{diagonalized}), a generic variational equation can be written as:
\begin{eqnarray}
\delta \xi^{(t+1)} &=& \left[Df(s)+\gamma Dh(s) \right]\cdot\delta \xi^t \;\;\;\;\;\; \text{mod(1)}.
\label{generic}
\end{eqnarray}
The largest Lyapunov exponent of Eq. (\ref{generic}) yields the MSF $\Psi_{\Omega,K}(\gamma)$ for a network topology specified by the parameter $\gamma$ containing Circle maps defined by $\Omega$ and $K$. For two maps coupled as shown in Fig. \ref{figure_1}(a), the coupling matrix ${\bf G}$ is written as:
\begin{equation}
{\bf G}=
 \begin{pmatrix}
 -1 & 1 \\
 1 & -1
\end{pmatrix},
\label{G}
\end{equation}
with eigenvalues $\lambda_1=0$ and $\lambda_2=-2$. The eigenvalue $\lambda_1=0$ does not influence the transversal stability of completely synchronized states since it corresponds to an eigenvector tangent to it.  

\begin{figure}[!htp]
\centering
\includegraphics[width=8.5cm,height=4.5cm]{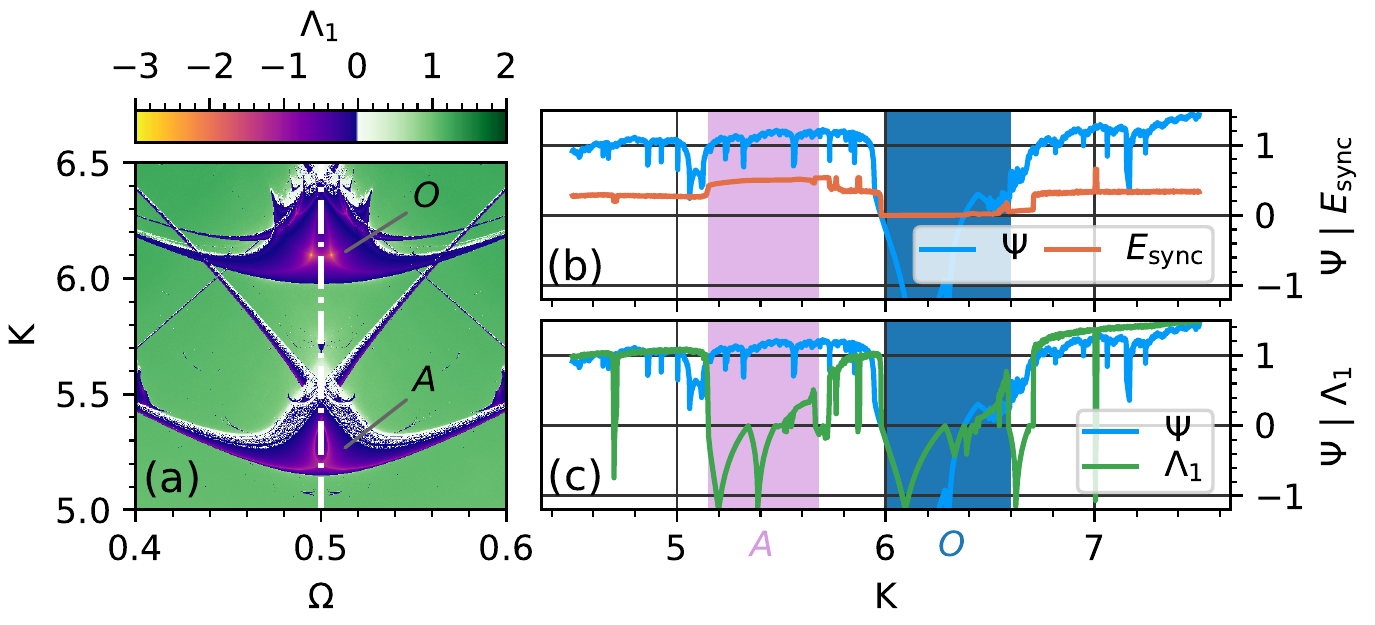}
\caption{For the system described in Eq. (\ref{Circle2}). (a) The largest Lyapunov exponent $\Lambda_1$ (colorcode) as a function of the control parameters $\Omega$ and $K$. The original window is identified by $O$ and the replica is identified by the letter $A$. (b) For $\Omega=0.5$ and $K \in [4.5,7.0]$ (dot-dashed line in (a)), the red curve stands for the synchronization error $E_\text{sync}$, while blue curve indicates $\Psi_{\Omega,K}(\gamma)$. (c) The green curve stands for $\Lambda_1$ and the blue curve $\Psi_{\Omega,K}(\gamma)$. The network parameters are $\sigma=0.3$ and $N=2$.}
\label{figure_7}
\end{figure}

Now, we first demonstrate the occurrence of replication of periodic windows for the coupling scheme adopted in Eq. (\ref{Circle2}) and for ${\bf G}$ prescribed by the matrix in Eq. (\ref{G}). By computing the largest nonzero Lyapunov exponent $\Lambda_1$, for $\sigma=0.3$, we show in Fig. \ref{figure_7}(a) the onset of one replica window marked with the letter $A$. Next, we fix the parameter $\Omega=0.5$ and analyze the transversal stability via MSF as a function of $K$ in the interval [$5.0$,$6.5$], i.e., the dashed line in Fig. \ref{figure_7}(a) crossing the original window $O$ and the replica $A$. Hence, in Fig. \ref{figure_7}(b), we validate our estimative of $\Psi_{\Omega,K}(\gamma)$ by comparing it with the synchronization error $E_\text{sync}$ defined in Eq. (\ref{sync_error}). In this figure, we observe that $\Psi_{\Omega,K}(\gamma)$ is successfully describing the stability of the synchronized state, i.e., $\Psi_{\Omega,K}(\gamma)<0$ for $E_\text{sync}=0$. Hence, in Fig. \ref{figure_7}(c), we demonstrate that, in fact, the synchronized state is transversaly unstable $\Psi_{\Omega,K}(\gamma)>0$ in the parameter region corresponding to the replica window with $\Lambda_1<0$, red in Fig. \ref{figure_7}(c).

Therefore, once the completely synchronized solution is unstable, the network approaches the phase cluster configuration allowing the out-of-phase behavior of the maps close of a stable periodic orbit composing the solutions observed for the replica windows. 

\section{Conclusions}

In summary, we investigate the role of the coupling in establishing regularity in the state-space of a networked system of discrete-time units. We observe the onset of stable periodic orbits occurring via replications of periodic windows in the parameter space of these systems. This phenomenon implies a significant gain to the stability domain of the network, enabling not only suppression of chaos for large regions of the system parameter space, but also increasing the system robustness to external perturbations such as localized shocks or noise. We characterize the dependence of the reported phenomenon regarding the main network parameters, i.e., network size and coupling intensity. We found the existence of optimal values of these parameters for the replications to occur. In particular, we emphasize that the replication phenomenon is very well observed in small networks. For networks with a high number of units, it may be difficult to visualize this phenomenon. In higher dimensions, the possibility of solutions with a larger number of positive Lyapunov exponents increases. For this case, the periodic windows are not well defined. Moreover, for larger networks, the trapping of trajectories in nonattracting chaotic sets dwelling inside periodic windows desynchronizes the network for times that are indefinitely long. 

Moreover, we report that the replication of periodic windows occurring in networked systems is closely related to the phenomenon of cluster synchronization networks. More specifically, since our network is composed of identical units, the occurrence of complete synchronization reduces the state-space dynamics to the uncoupled behavior. Consequently, there would not be the onset of new stable periodic orbits in the system. However, we verify that the network solution in the replica windows is cluster synchronized. The units are completely synchronized within clusters but out of phase from one cluster to another. These phase clusters spontaneously generate the internal disturbances to replicate the periodic windows of the adopted map. We show that, in a peculiar case in which the number of phase clusters equals the network size, traveling waves arise in the network state-space where all units are out-of-phase around the same periodic orbit. Finally, we employ the master stability function analysis to demonstrate that the occurrence of replica windows requires the completely synchronized states to be transversally unstable. 

\acknowledgments
This work was supported by the Deutsche Forschungsgemeinschaft (DFG) via the project number 163436311-SFB-910. E.S.M acknowledges the support by the Deutsche Forschungsgemeinschaft (DFG) via the project number 454054251. R.O.M acknowledges the support by FAPESP via the project number 15/50122-0.


\end{document}